\documentclass[twocolumn,aps,prd,longbibliography]{revtex4-1}
\usepackage{graphicx}  
\usepackage{dcolumn}   
\usepackage{bm}        
\usepackage{verbatim}   
\usepackage{graphicx}
\usepackage{epstopdf}
\usepackage{psfrag}
\usepackage{footnote}
\usepackage{epsfig}
\usepackage{subfigure}
\usepackage{amssymb}
\usepackage{xcolor}
\usepackage{amsmath}
\usepackage[most]{tcolorbox}
\usepackage{lipsum} 
\usepackage{color}
\usepackage[colorlinks=true,linkcolor=blue,allcolors=blue]{hyperref}%

\begin{document}

	
	
\headsep = 40pt
\title{Dynamic Modulation Yields One-Way Beam Splitting}
\author{Sajjad Taravati$^{1,2}$ and Ahmed A. Kishk$^{1}$}
\affiliation{$^{1}$Department of Electrical and Computer Engineering, Concordia University, Montr\'{e}al, Quebec {H3G 2W1}, Canada\\
	$^{2}$Department of Electrical and Computer Engineering, University of Toronto, Toronto, Ontario {M5S 2E4}, Canada\\	
	email: sajjad.taravati@utoronto.ca}
\date{\today}	

\begin{abstract}
	This article demonstrates the realization of an extraordinary beam splitter, exhibiting one-way beam splitting-amplification. Such a dynamic beam splitter operates based on nonreciprocal and synchronized photonic transitions in obliquely illuminated space-time-modulated (STM) slabs which impart the coherent temporal frequency and spatial frequency shifts. As a consequence of such unusual photonic transitions, a is exhibited by the STM slab. Beam splitting is a vital operation for various communication systems, including circuit quantum electrodynamics, and signal-multiplexing and demultiplexhg. Despite the beam splitting is conceptually a simple operation, the performance characteristics of beam splitters significantly influence the repeatability and accuracy of the entire system. As of today, there has been no approach exhibiting a nonreciprocal beam splitting accompanied with transmission gain and an arbitrary splitting angle. Here, we show that oblique illumination of a periodic and semi-coherent dynamically-modulated slab results in coherent photonic transitions between the incident light beam and its counterpart space-time harmonic (STH). Such transitions introduce a unidirectional synchronization and momentum exchange between two STHs with same temporal frequencies, but opposite spatial frequencies. Such a beam splitting technique offers high isolation, transmission gain and zero beam tilting, and is expected to drastically decrease the resource and isolation requirements in communication systems. In addition to the analytical solution, we provide a closed-form solution for the electromagnetic fields in STM structures, and accordingly, investigate the properties of the wave isolation and amplification in subluminal, superluminal and luminal ST modulations.
\end{abstract}

\maketitle
\section{Introduction}
Beam splitters are quintessential elements of communication systems~\cite{schneider2014chip,hammer2015microwave,tan2017planar,watanabe1980novel,zhu2006microwave,hwang2012spatial,cooper2012novel,watanabe1980novel}. In the microwave regime, beam splitters are required for the generation of single photons in the circuit quantum electrodynamics~\cite{wallraff2004strong,blais2004cavity,mariantoni2008two,schneider2014chip,houck2007generating,bozyigit2011correlation,bozyigit2011antibunching}, heterodyne mixer arrays~\cite{tan2017planar}, and wave engineering and signal-multiplexing and demultiplexhg~\cite{watanabe1980novel,zhu2006microwave,hwang2012spatial,cooper2012novel,watanabe1980novel}. However, the realization of microwave on-chip beam splitters is still under research and development~\cite{mariantoni2010planck,schneider2014chip,hammer2015microwave}. In spite of the immense scientific attempts for the realization of efficient beam splitters, beam splitters are restricted to reciprocal response and suffer from substantial transmission loss. As a consequence, the resource requirements of the overall system, including demand for high power microwave sources and isolators, will be increased. 

This paper presents the application of space-time-modulated (STM) structures to extraordinary beam splitting. As of today, various applications of STM structures have been reported, where normal incidence of the light beam to the STM structure yields unusual interaction with electromagnetic wave~\cite{Cassedy_PIEEE_1967,Fan_NPH_2009,Taravati_PRAp_2018,Taravati_PRB_2017,salary2018time,salary2018electrically,Taravati_Kishk_TAP_2018}. These applications include but not limited to the parametric traveling-wave amplifiers~\cite{Cullen_NAT_1958,Tien_JAP_1958,tien1958traveling,Oliner_PIEEE_1963}, isolators~\cite{wentz1966nonreciprocal,bhandare2005novel,Fan_PRL_109_2012,Taravati_PRB_2017,Chamanara_PRB_2017,Taravati_PRB_SB_2017,kim2017complete,sohn2018time}, metasurfaces~\cite{Alu_PRB_2015,Fan_APL_2016,Fan_mats_2017,wu2018transparent}, pure frequency mixer~\cite{Taravati_PRB_Mixer_2018}, circulators~\cite{Wang_TMTT_2014,estep2014magnetic,reiskarimian2018integrated}, and mixer-duplexer-antenna system~\cite{Taravati_APS_2015,Taravati_LWA_2017}. Nevertheless, there has been a lack of investigation on the properties of STM media under oblique incidence and its applications.

Here, we introduce a one-way, tunable and highly efficient beam splitter and amplifier based on coherent photonic transitions through the oblique illumination of STM structures. The contributions of this paper are as follows.

1) In contrast to conventional beam splitters which are restricted to reciprocal response with more than 3 dB insertion loss, the proposed STM beam splitter is capable of providing nonreciprocal response with transmission gain. It can be also used in antenna applications, where the transmitted and received waves are engineered appropriately.

2) We show that the STM beam splitter presents an efficient performance for both collimated and non-collimated incidence beam with no output beam tilt. This is very interesting as conventional passive beam splitters suffer from poor performance for non-collimated beams and provide an undesired output beam tilt. 

3) It is demonstrated that the angle of transmission and the amplitude of the transmitted beams depend on the ST modulation parameters. Hence, the ST modulation parameters provide the leverage for achieving the desired angle of transmission for the two output beams of the STM beam splitter. In addition, unequal power division between the output beams can be achieved by varying the ST modulation parameters.

4) Here, we present the first application of obliquely illuminated STM slabs. Consequently, for the first time, the scheme and results for the finite difference time-domain (FDTD) simulation results for oblique incidence to a STM slab at microwave frequencies is presented.

5) A closed-form solution is presented that provides a deep insight into the wave propagation inside the STM beam splitters and the difference between the subluminal, luminal and superluminal ST modulations.

6) The analysis of the STM beam splitter is further accomplished by investigation of its analytical three dimensional dispersion diagrams, achieved by Bloch-Floquet decomposition of space-time harmonics (STHs).

Accordingly, the rest of the paper is structured as follows. Section~\ref{sec:oper} presents the operation principle of the proposed STM beam splitter. In Sec.~\ref{sec:analyt}, we derive the analytical solution for oblique electromagnetic wave propagation inside the STM beam splitter based on the Bloch-Floquet representation of the electromagnetic fields. Then, Sec.~\ref{sec:num} presents the time and frequency domains numerical simulation results for the beam splitting and amplification in the STM beam splitter. Next, the closed form solution will be provided in Sec.~\ref{sec:closed}, which gives a leverage for understanding the wave propagation and transitions in STM structures. A short discussion on practical realization of superluminal STM structures at different frequencies will be presented in Sec.~\ref{sec:prac}. Finally, Sec.~\ref{sec:conc} concludes the paper.

\section{Operation Principle}\label{sec:oper}
\begin{figure}
	\begin{center}
		\includegraphics[width=1\linewidth]{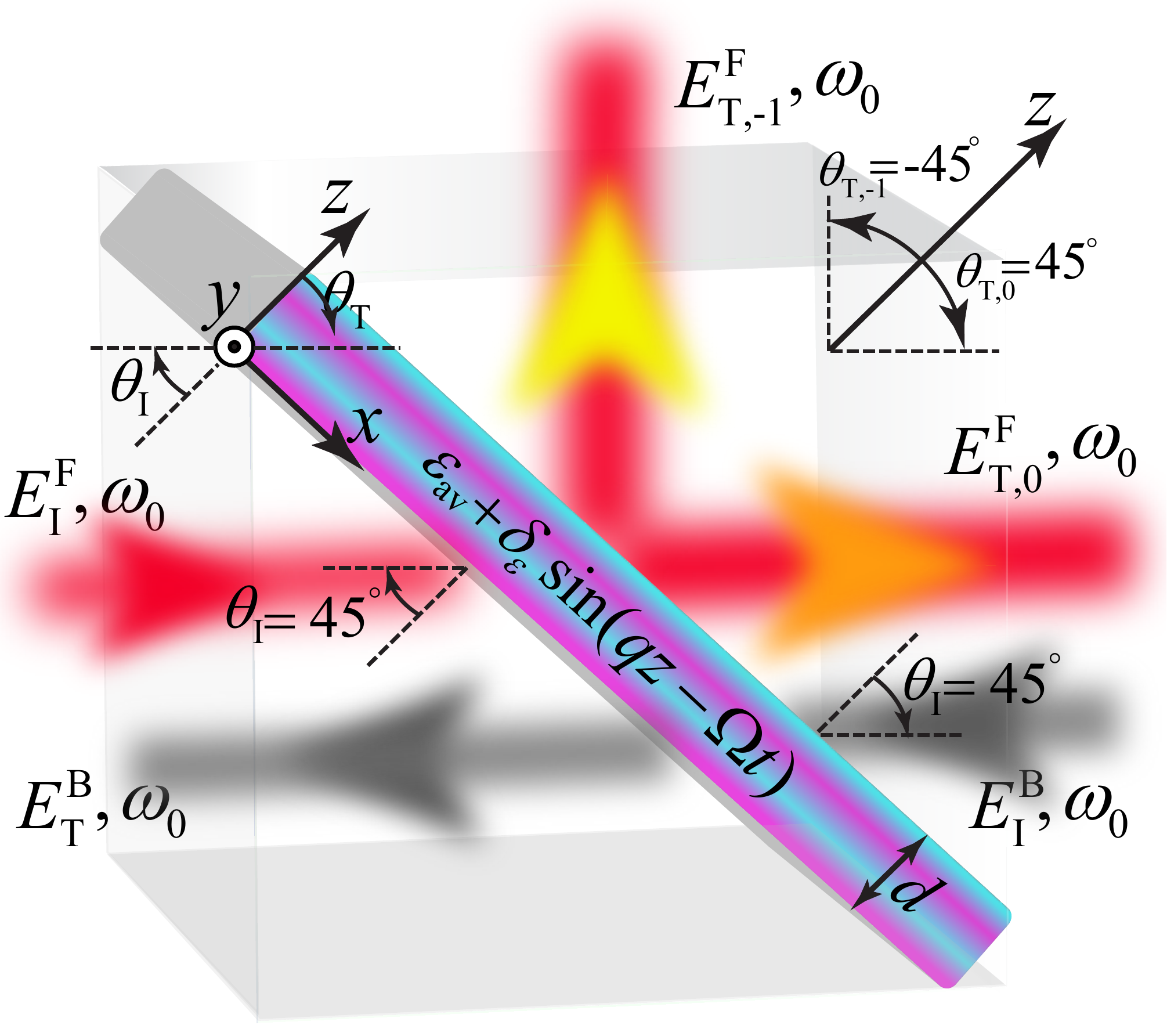}
		\caption{Schematic of nonreciprocal beam splitting in a STM slab. The slab varies in time two times faster than the input wave.}
		\label{Fig:bs_sch}
	\end{center}
\end{figure}

Figure~\ref{Fig:bs_sch} sketches the nonreciprocal beam transmission and splitting in a STM slab. By appropriate design of the band structure, that is, the ST modulation format and its associated temporal and spatial modulation frequencies, unidirectional energy and momentum exchange between the incident wave-under angle of incidence and transmission $\theta_{\text{I}}=\theta_{\text{T},0}=45^\circ$ and temporal frequency $\omega_0$- to the first lower STH-under angle of transmission $\theta_{\text{T},-1}=-45^\circ$ and temporal frequency $\omega_0$- will occur. Assuming $\text{TM}_{y}$ or $E_y$ polarization, the electric field of the incident light beam in the forward $+z$-direction may be expressed as
\begin{equation}\label{eqa:elec}
E_\text{I}^\text{F}(x,z,t)=E_{0} e^{-i \left[k_x x+k_z z -\omega_0 t \right]},
\end{equation}
is traveling in the $+z$-direction under the angle of incidence $\theta_{\text{I}}=45^\circ$ and impinges to the periodic STM slab. The $x$- and $z$-components of the spatial frequency read $k_x=k_0 \sin(\theta_{\text{I}})$ and $k_z=k_0 \cos(\theta_{\text{I}})$, respectively, in which $k_0=\omega_0/v_\text{b}=\omega_0\sqrt{\epsilon_\text{r}}/c$, with $\omega_0$ being the temporal frequency of the incident wave, $v_\text{b}$ denoting the phase velocity in the background medium, $\epsilon_\text{r}$ representing the relative electric permittivity of the background medium, and $c$ denoting the speed of light in vacuum.
The STM slab assumes a sinusoidal ST-varying permittivity, as
\begin{equation}\label{eqa:permit}
	\epsilon(z,t)=\epsilon_\text{av} + \delta_{\epsilon} \sin(q z-\Omega t),
\end{equation}
where $\epsilon_\text{av}=\epsilon_\text{r} +\delta_{\epsilon}$ is the average permittivity of the slab, $\delta_{\epsilon}$ denotes the modulation strength, $\Omega=2 \omega_0$ is the modulation temporal frequency, and 
\begin{equation}
q= \frac{2 k_0}{\gamma},
\end{equation} 
represents the spatial frequency of the modulation, with $\gamma=v_\text{m}/v_\text{b}$ being the ST velocity ratio, where $v_\text{m}$ and $v_\text{b}$ are, the phase velocity of the modulation and the background medium, respectively. Since the slab permittivity is periodic in space and time, with spatial frequency $q$ and temporal frequency $2\omega_0$, the electric field inside the slab may be decomposed into ST Bloch-Floquet waves as
\begin{subequations}\label{eq:field_ins}
\begin{equation}\label{eq:field_ins_E}
\mathbf{E}_\text{S}(x,z,t)=\mathbf{\hat{y}}\sum_{m=-M}^M  A_{m} e^{-i \left(k_x x+k_{z,m} z -\omega_m t \right)},
\end{equation}
and
\begin{equation}\label{eq:field_ins_H}
		\begin{split}
		&\mathbf{H}_\text{S}(x,z,t)=
		\frac{1}{\eta_\text{S}}\left[\mathbf{\hat{k}}_\text{S} \times \mathbf{E}_\text{S} (x,z,t)\right]\\
		&=\sum_{m =  - M}^M      \bigg[-\mathbf{\hat{x}} \frac{k_{z,m}}{\mu_0 \omega_m } + \mathbf{\hat{z}}  \frac{\sin(\theta_\text{I}) }{\eta_\text{S}} \bigg] A_{m} e^{-i \left(k_x x+k_{z,m} z -\omega_m t \right)} .
		\end{split}
\end{equation}
where $M\rightarrow\infty$ is the number of STHs. In Eq.~\eqref{eq:field_ins}, $\eta_\text{S}=\sqrt{\mu_0/(\epsilon_0\epsilon_r)}$, and $A_m$ represents the unknown amplitude of the $m$th STH, characterized by the spatial frequency
\begin{equation}
k_{z,m}=\beta_{0}+mq, 
\end{equation}
 and the temporal frequency
\begin{equation} 
\omega_m=\omega_0+m\Omega=(1+2m) \omega_0,
\end{equation}  
\end{subequations}  
with $\beta_{0}$ being the unknown spatial frequency of the fundamental harmonic. The unknowns of the electric field, that is, $A_m$ and $\beta_0$, will be found through satisfying Maxwell's equations.
	
The transmission angle of the $m$th transmitted STH, $\theta_{\text{T},m}$, satisfies the Helmholtz relation as
\begin{equation}\label{eqa:helm}
	k_{0}^2 \sin^2(\theta_{\text{I}})+k_{m}^2 \cos^2(\theta_{\text{T},m})=k_{m}^2,
\end{equation}
where $k_{m}=\omega_m/v_\text{b}$ denotes the wavenumber of the $m$th transmitted STH outside the STM slab. Solving Eq.~\eqref{eqa:helm} for $\theta_{\text{T},m}$ yields
\begin{subequations}
\begin{equation}\label{eqa:refl_trans_angl}
\begin{split}
	\theta_{\text{T},m}=\sin^{-1} \left(\frac{k_x}{k_m} \right)&=\sin^{-1} \left(\frac{\sin(\theta_\text{I})}{1+m \Omega/\omega_0} \right)\\
	&=\sin^{-1} \left(\frac{\sin(\theta_\text{I})}{1+2m} \right).
\end{split}
\end{equation} 

Equation~\eqref{eqa:refl_trans_angl} demonstrates the spectral decomposition of the transmitted wave. Consequently, the fundamental STH, $m=0$, and the first lower STH, $m=-1$, with equal temporal frequency $\omega_0$, will be respectively transmitted under the angles of transmission of
\begin{equation}
\begin{split}
\theta_{\text{T},0}&=\theta_{\text{I}}=45^\circ, \\ \theta_{\text{T},-1}&=-\theta_{\text{I}}=-45^\circ.
\end{split}
\end{equation} 
\end{subequations}
so that they are transmitted under $90^\circ$ angle difference, presenting the desired beam splitting. The scattering angle of the $m$th STH inside the STM slab reads
\begin{equation} 
\theta_{\text{S},m}=\tan^{-1} \left( \frac{k_x}{k_{z,m} }\right).
\end{equation}

In addition, the transmitted electric field from the slab may be found as 
\begin{equation} 
\begin{split}
\mathbf{E}_{\text{T}}(x,z,t)&=\mathbf{E}_\text{S}(x,z,t) e^{-i k_{z,m}  z}\\
&=\mathbf{\hat{y}}\sum_{m=-M}^M  A_{m} e^{-i \left(k_x x+k_{z,m} [d+z] -\omega_m t \right)} .
\end{split}
\end{equation}

The sourceless wave equation reads
\begin{equation}
\nabla^2 \mathbf{E}_\text{S}(x,z,t) - \frac{1}{{{c^2}}}\frac{{{\partial ^2} \left[\epsilon (z,t)\mathbf{E}_\text{S}(x,z,t) \right]}}{{\partial {t^2}}}=0.
\label{eqa:wave_eq}
\end{equation}  
Substituting Eqs.~\eqref{eqa:permit} and~\eqref{eq:field_ins} into Maxwell's equations, yields a matrix equation as
\begin{subequations}
\begin{equation}\label{eqa:eq}
	[K] \vec{A}=0,
\end{equation}
where $[K]$ is a $(2M+1)\times(2M+1)$ matrix with elements 
\begin{equation}
\begin{split}
&K_{m,m} = \epsilon_\text{av} -\frac{k_x ^2+ k_{z,m}^2}{k_0^2},\\
&K_{m,m-1}=i\frac{\delta_\epsilon}{2},\\
&K_{m,m+1}=-i\frac{\delta_\epsilon}{2},
\end{split}
\end{equation}
\end{subequations}
and where $\vec{A}$ represents a $(2M+1)\times 1$ vector containing $A_m$ coefficients. Equation~\eqref{eqa:eq} has nontrivial solution if
\begin{equation}\label{eqa:disp}
\text{det}\left\{[K]\right\}=0.
\end{equation}	 

Equation~\eqref{eqa:disp} represents the dispersion relation of the STM beam splitter which provides the unknown spatial frequency of the fundamental ST harmonic for a given frequency, i.e., $\beta_0(\omega_0)$. After finding the $\beta_0(\omega_0)$, the $[K]$ matrix in Eq.~\eqref{eqa:eq} is known and therefor, the unknown amplitude of the STHs $A_m$ will be calculated using Eq.~\eqref{eqa:eq}.

\section{Analytical 3D Dispersion Diagram}\label{sec:analyt}
\begin{figure}
	\begin{center}
		\includegraphics[width=0.99\linewidth]{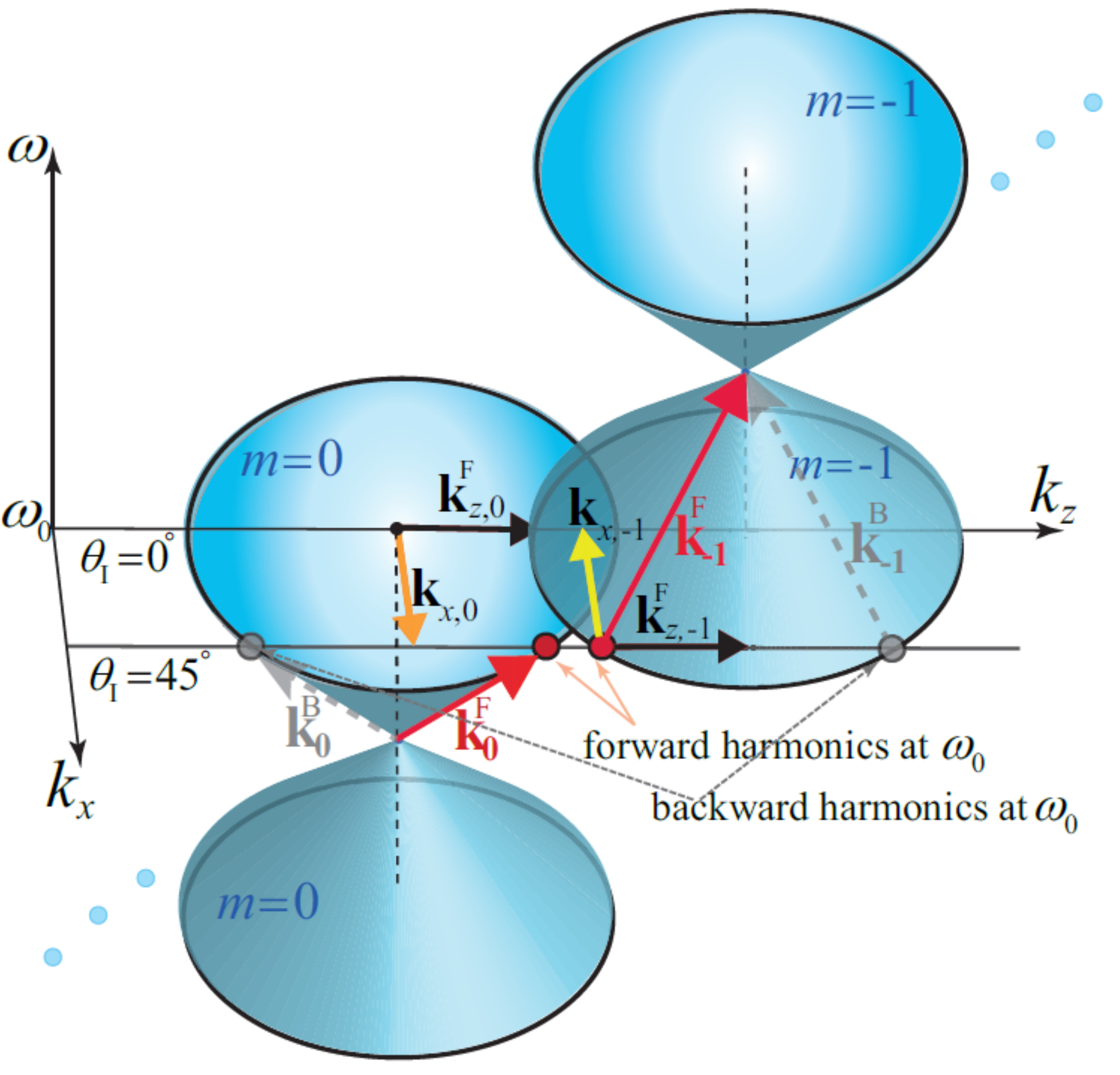}
		\caption{Qualitative representation of the periodic three dimensional dispersion diagram for the periodic STM slab in Fig.~\ref{Fig:bs_sch}. The medium is under oblique illumination of $\theta_{\text{I}}=45^\circ$ at the fundamental harmonic $m=0$, corresponding to the temporal frequency $\omega_0$, where $\textbf{k}_{x,0}=\mathbf{\hat{x}} k_x=\mathbf{\hat{x}} k_0 \sin(\theta_{\text{I}})$. The lower STH, $m=-1$, provides the same temporal frequency as the fundamental harmonic, $|\omega_m|=|\omega_0(1+2m)|_{m=-1}=\omega_0$, but opposite $x$-component of the spatial frequency, that is, $\textbf{k}_{x,-1}=-\mathbf{\hat{x}} k_x=-\mathbf{\hat{x}} k_0 \sin(\theta_{\text{I}})$.}
		\label{Fig:3D_disp_sch}
	\end{center}
\end{figure}

Figure~\ref{Fig:3D_disp_sch} presents a qualitative illustration of the three dimensional dispersion diagram in the STM medium in Fig.~\ref{Fig:bs_sch} achieved using Eq.~\eqref{eqa:disp}. This diagram is formed by $2M+1$ periodic set of double cones (here, only $m=0$ and $m=-1$ harmonics are shown), each of which representing a STH, with apexes at $k_x = 0$, $k_z = -m q$ and $\omega=-2m \omega_0$, and the slope of $v_m$ with respect to $k_z-k_x$ plane. Consider oblique incidence of a wave, representing the fundamental harmonic $m=0$ with temporal frequency $\omega_0$, propagating along [$+x$,$+z$] direction. It is characterized by $x$- and $z$-components of the spatial frequency, $\textbf{k}_{x}=\mathbf{\hat{x}} k_x$ and $\textbf{k}_{z}^\text{F}=\mathbf{\hat{z}} k_z$. The wave impinges to the medium under the angle of incidence $\theta_\text{I}=45^\circ$ and excites an infinite number of (we truncate it to $2M+1$) STH waves, with different spatial and temporal frequencies of $[k_x,k_{z,m}]$ and $\omega_{m}$. However, interestingly, the first lower STH $m=-1$ offers similar characteristics as the fundamental harmonic, that is, the identical temporal frequency of $\omega_{0}$ and identical $z$-component of the spatial frequency of $\textbf{k}_{z,-1}^\text{F}=\textbf{k}_{z,0}^\text{F}$, but opposite $x$-component of the spatial frequency of  $\textbf{k}_{x,-1}=-\textbf{k}_{x,0}$. Hence, $m=-1$ harmonic propagates along [$-x$,$+z$] direction. In general the $x$-component of the $m$th STH reads  $\textbf{k}_{x,m}=-\textbf{k}_{x,-m-1}$). Moreover, since $\omega_{m}=\omega_{-m-1}$, the undesired STHs acquire temporal frequency of $2m\omega_0$, and far away from the fundamental harmonic. Thus, most of the incident energy is residing in $m=0$ and $m=-1$ harmonics, both at $\omega_{0}$, respectively transmitted under $\theta_\text{T,0}=\theta_\text{I}$ and $\theta_\text{T,-1}=-\theta_\text{I}$ transmission angles with $2\theta_\text{I}$ angle difference.

The exchange of the energy and momentum between the fundamental and first lower harmonic occurs only for the forward, $+z$, wave incidence. This may be observed from Fig.~\ref{Fig:3D_disp_sch}, as the forward harmonics (red circles, where $\partial \omega/\partial k_z>0$) are very close, whereas the backward harmonics (grey circles, where $\partial \omega/\partial k_z<0$) are far apart from each other. Therefore, a nonreciprocal transition of energy is achieved from the incident wave under $\theta_\text{I}=45^\circ$ to the first STH under $\theta_\text{T,-1}=-45^\circ$, through the ST modulation under $\theta_\text{mod}=0^\circ$. 

\begin{figure}[h!]
	\begin{center}
		\subfigure[]{\label{Fig:disp-a}
			\includegraphics[width=0.95\linewidth]{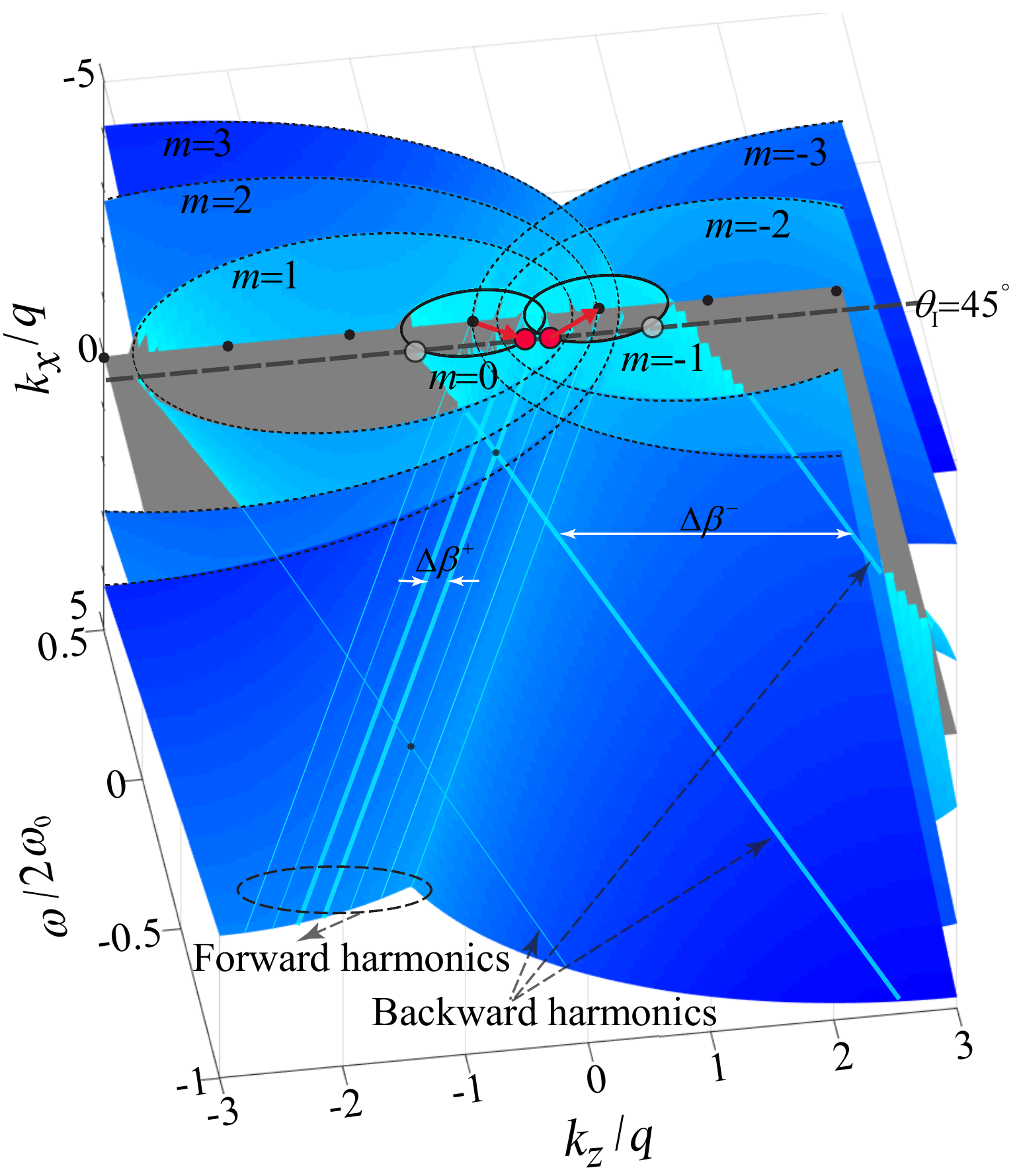}  }
		\subfigure[]{\label{Fig:disp-b}
			\includegraphics[width=0.95\linewidth]{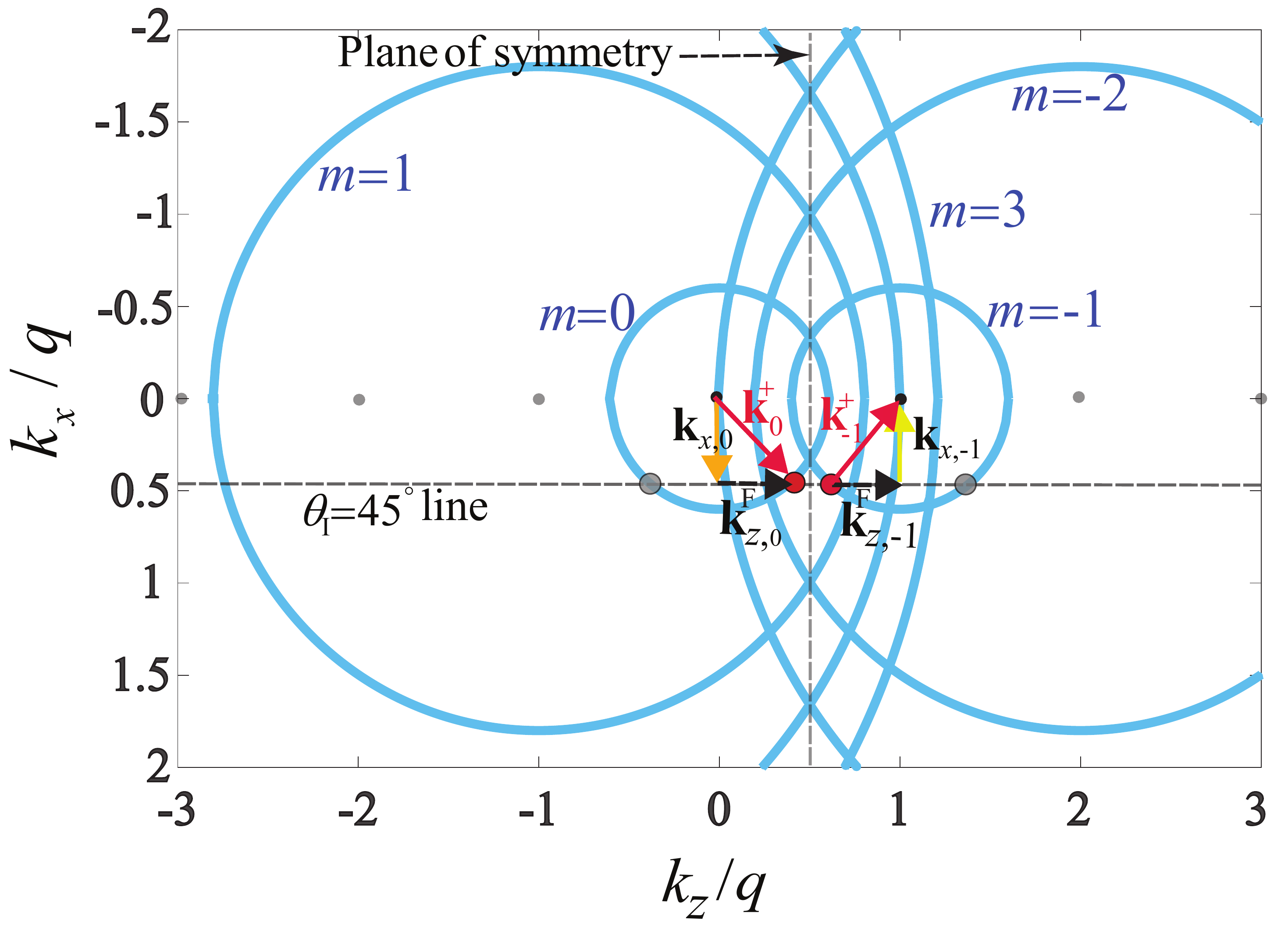}  }
		\caption{Analytical dispersion diagram of the periodic STM slab in Fig.~\ref{Fig:bs_sch} for $\gamma=1.2$ computed using Eq.~\eqref{eqa:disp}. The forward incidence under $\theta_{\text{I}}=45^\circ$ corresponding to $k_x/q=0.4243$ excites the $m=-1$ STH, resulting in a strong exchange of energy between  $m=0$ and $m=-1$ harmonics, with the identical temporal frequency of $\Omega_0$. (a)~Three dimensional dispersion diagram constituted of an array of periodic cones~\cite{Taravati_PRB_2017}. (b)~Isofrequency diagram at $\omega=\omega_0$ presents an infinite set of circles centered at $(k_z/q,k_x/q)=(-m,0)$ with radius $\gamma(0.5+m)$.}
		\label{Fig:}
	\end{center}
\end{figure}

Figure~\ref{Fig:disp-a} shows the analytical solution for three dimensional dispersion diagram of the STM medium in Fig.~\ref{Fig:bs_sch}, computed using Eq.~\eqref{eqa:disp} for $\gamma=1.2$.  For a given frequency, this three dimensional diagram provides the two dimensional $k_z/q-k_x/q$ isofrequency diagram of the medium. Figure~\ref{Fig:disp-b} plots the isofrequency diagram at $\omega/2\omega_0=0.5$ (or $\omega=\omega_0$), containing an infinite periodic set of circles centered at $(k_z/q,k_x/q) = (-m,0)$ with radius $\gamma(0.5+m)$.

It may be seen from Figs.~\ref{Fig:disp-a} and~\ref{Fig:disp-b} that at $\omega=\omega_0$, the $m=0$ and $m=-1$ STHs offer identical isofrequency circles. However, their associated forward harmonics (red circles) are very close to each other whereas their associated backward harmonics (grey circles) are significantly separated. For a nonzero velocity ratio ($\gamma>0$), the forward and backward STHs acquire different distances, i.e. $\Delta\beta^\pm=k_{z,m+1}^\pm-k_{z,m}^\pm$~\cite{Taravati_PRB_2017}. Particularly, as $\gamma$ increases, $\Delta\beta^-$ increases and $\Delta\beta^+$ decreases. As a result, at the limit of $\gamma=1$ the forward harmonics acquire distances $\Delta\beta^+/q=0$, and the backward harmonics acquire distances $\Delta\beta^-/q=2$. Hence, increasing $\gamma$ results in the significant enhancement in the nonreciprocity of the medium, so that the forward harmonic waves tend to merge together ($\Delta\beta^+\rightarrow0$) and exchange their energy and momentum, whereas the backward harmonics tend to separate from each other ($\Delta\beta^+\rightarrow2$) (Fig.~\ref{Fig:disp-a}). Hence, such a dynamic modulation has nearly no effect on the backward incident beam.  
%
	
%

%
\section{Numerical Simulation Results}\label{sec:num}
\begin{figure}[h!]
	\begin{center}
		\includegraphics[width=0.85\columnwidth]{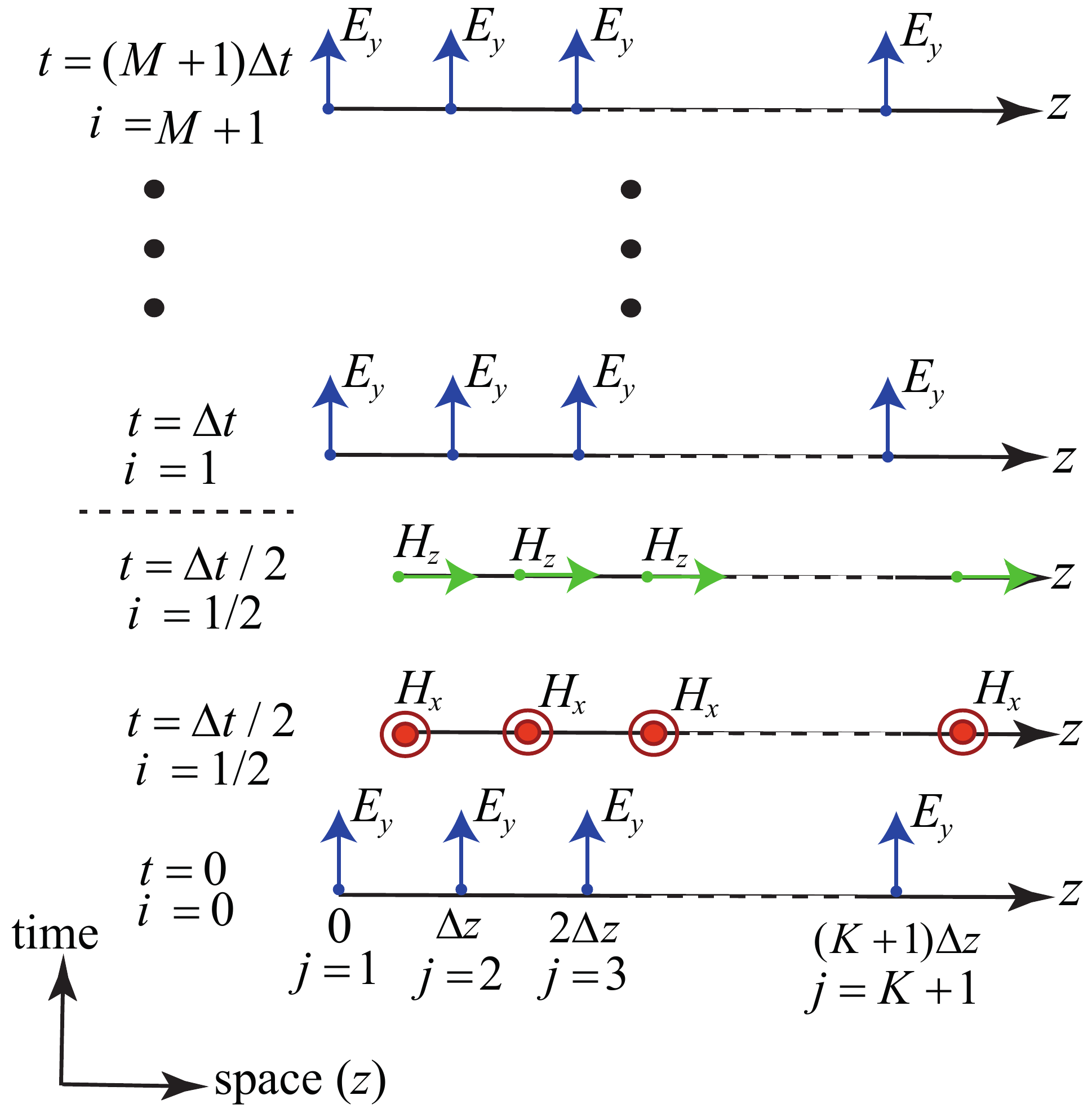} 
		\caption{General representation of the finite-difference time-domain scheme for numerical simulation of the oblique incidence of an $E_y$ wave to STM beam splitter.} 
		\label{Fig:FDTD}
	\end{center}
\end{figure}

We next verify the above theory by finite difference time-domain (FDTD) numerical simulation of the dynamic process through solving Maxwell's equations. Figure~\ref{Fig:FDTD} plots the implemented finite-difference time-domain scheme for numerical simulation of the oblique wave impinging to the STM beam splitter. We first discretize the medium to $K+1$ spatial samples and $M+1$ temporal samples, with the steps of $\Delta z$ and $\Delta t$, respectively.
Next, the finite-difference discretized form of the first two Maxwell's equations for the electric and magnetic fields (considering Eq.~\eqref{eq:field_ins}) will be simplified to
\begin{subequations}
	\begin{equation}\label{eqa:num_Max1c}
	\begin{split}
	H_x\lvert_{j+1/2}^{i+1/2}=&\left(1-\Delta t  \right)	H_x\lvert_{j+1/2}^{i-1/2}+\dfrac{\Delta t}{ \mu_0 \Delta z} \left( E_y\lvert_{j+1}^{i}-E_y\lvert_{j}^{i} \right) 
	\end{split}	
	\end{equation}
	\begin{equation}\label{eqa:num_Max1d}
	\begin{split}
	H_z\lvert_{j+1/2}^{i+1/2}=&\left(1-\Delta t  \right)	H_z\lvert_{j+1/2}^{i-1/2}-\dfrac{\Delta t}{ \mu_0 \Delta z} \left( E_y\lvert_{j+1}^{i}-E_y\lvert_{j}^{i} \right) 
	\end{split}	
	\end{equation}
	\begin{equation}\label{eqa:num_Max2c}
	\begin{split}
	E_y& \lvert_{j}^{i+1}= \left(1- \dfrac{\Delta t\epsilon'\lvert_{j}^{i}}{\epsilon \lvert_{j}^{i+1/2}}  \right) E_y \lvert_{j}^{i}+ \dfrac{\Delta t/\Delta z}{ \epsilon \lvert_{j}^{i+1/2}} \\
	&.\left[ \left( H_x \lvert_{j+1/2}^{i+1/2}-H_x\lvert_{j-1/2}^{i+1/2} \right) -\left( H_z \lvert_{j+1/2}^{i+1/2}-H_z\lvert_{j-1/2}^{i+1/2} \right)\right] 
	\end{split}	
	\end{equation}
\end{subequations}
where $\epsilon'=\partial \epsilon(z,t)/\partial t=-\Omega \delta_{\epsilon} \cos(qz-\Omega t)$.

\begin{figure}
	\centering
	\includegraphics[width=1\linewidth]{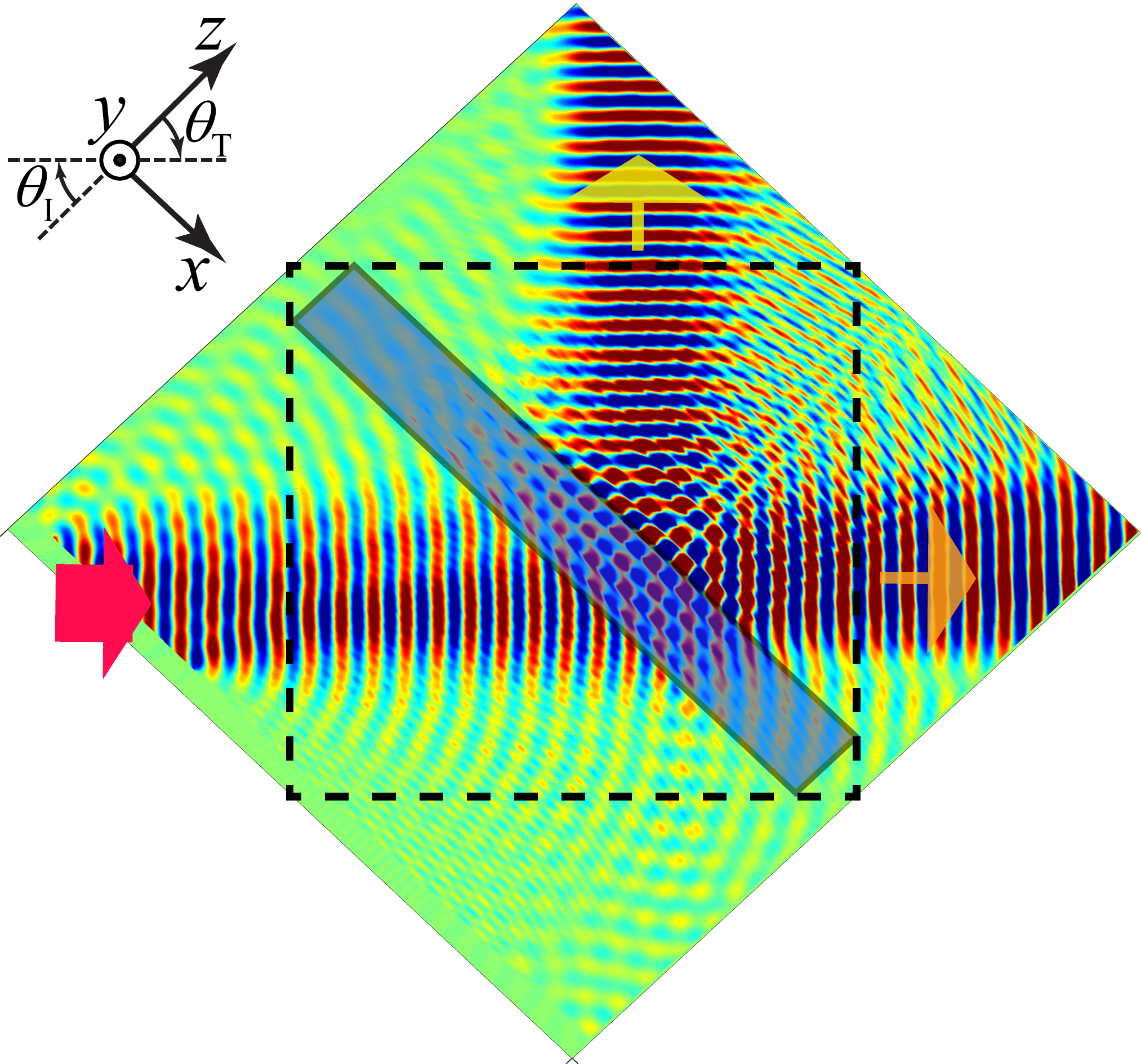}  
	\caption{Nonreciprocal beam splitting in periodically STM slab. FDTD numerical simulation for the forward wave incidence to the slab, from the left, with $\theta_\text{I}=45^\circ$.}
	\label{Fig:res_a}
\end{figure}

\begin{figure}
	\centering
	\subfigure[]{\label{Fig:res_c}
		\includegraphics[width=1\linewidth]{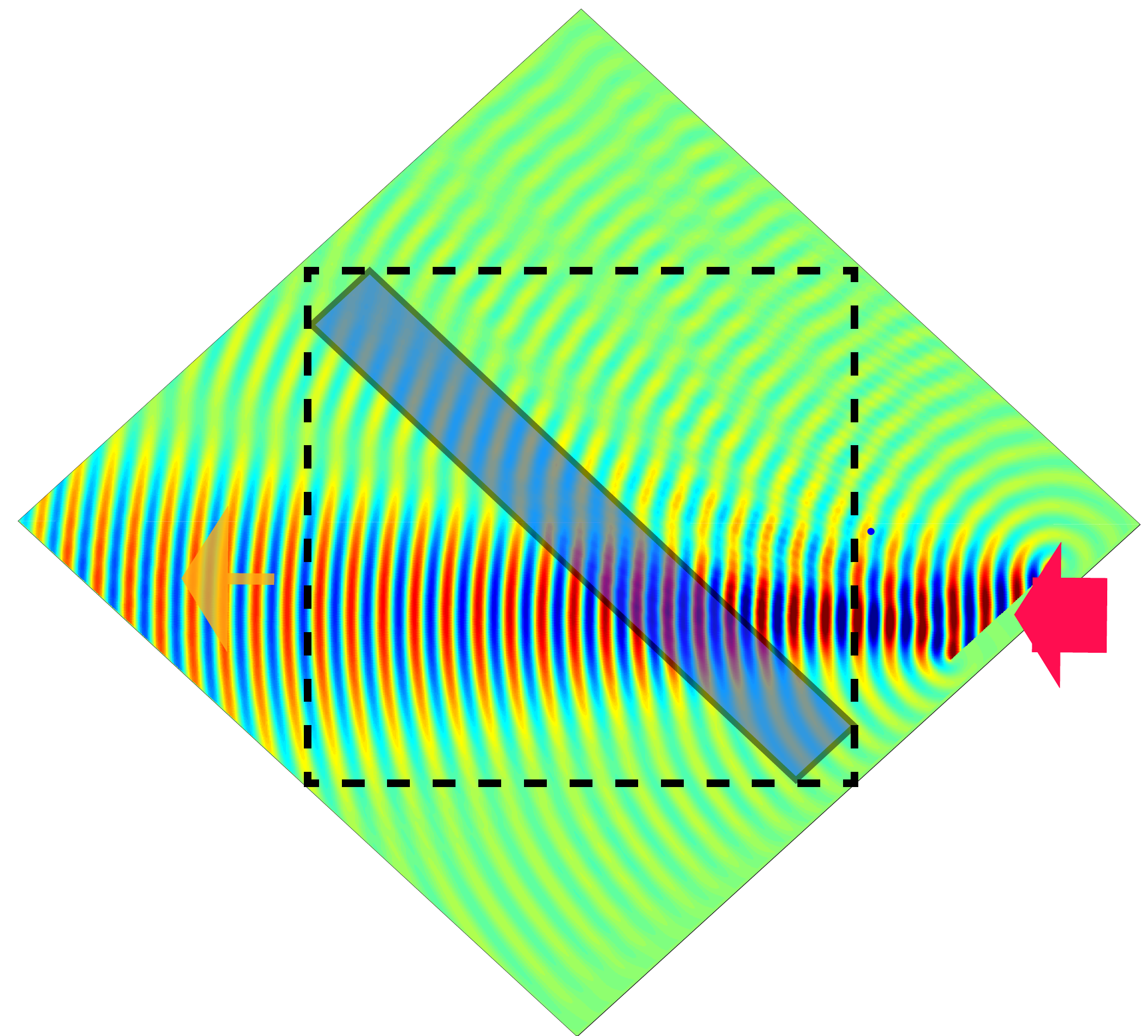}  }
	\subfigure[]{\label{Fig:res_d}
		\includegraphics[width=1\linewidth]{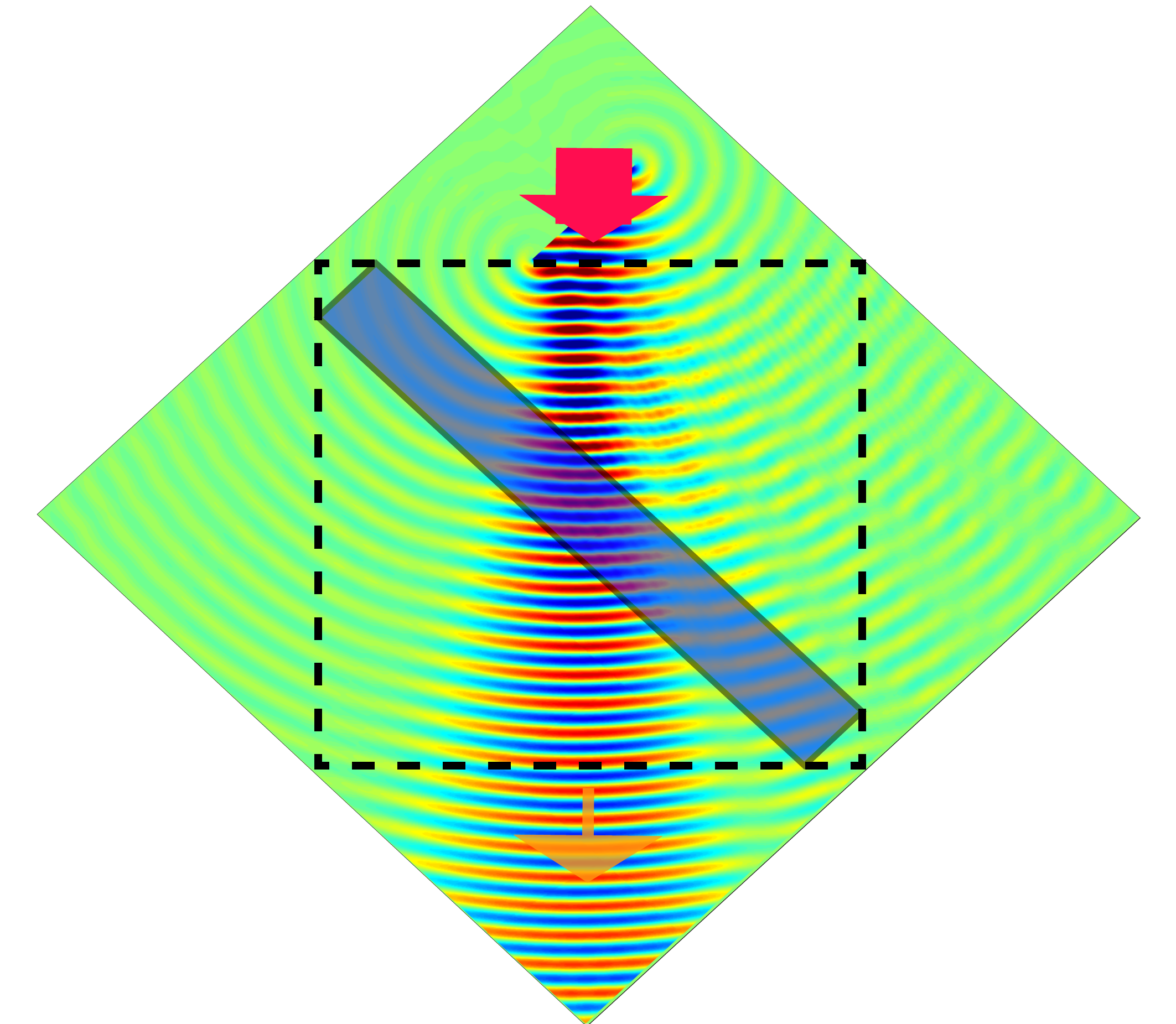}  }
	\caption{Nonreciprocal beam splitting in periodically STM slab. FDTD numerical simulation for the wave incidence to the slab, (a)~From the right with $\theta_\text{I}=45^\circ$. (b)~From the top, i.e., $\theta_\text{I}=-45^\circ$.}
	\label{Fig:RES2}
\end{figure}

\begin{figure}[h!]
	\begin{center}
			\includegraphics[width=0.95\linewidth]{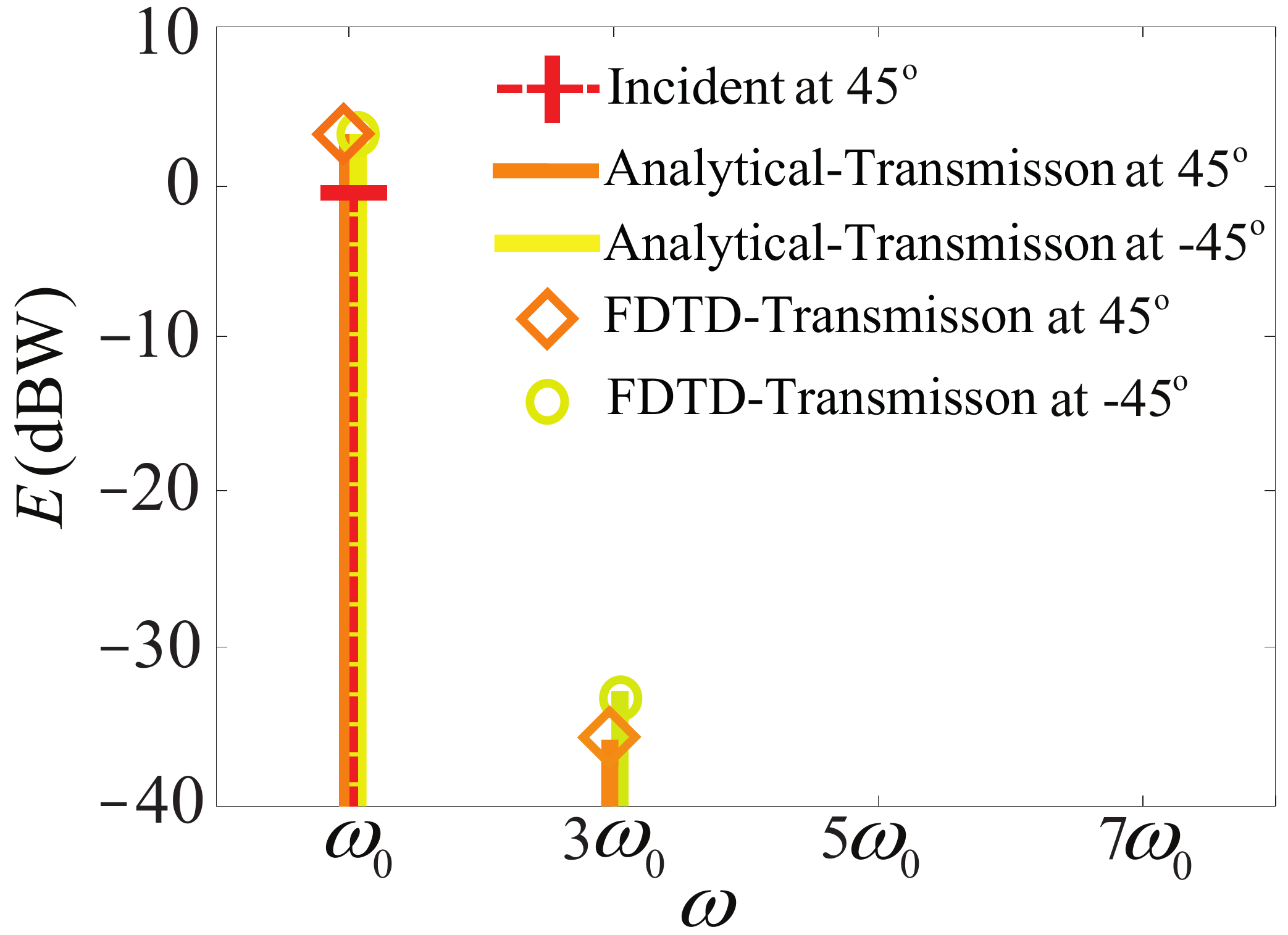}  
		\caption{Comparison of the analytical and numerical results for the frequency spectrum of the incident and transmitted electric fields in Fig.~\ref{Fig:res_a}, i.e., wave incidence to the slab from the left with $\theta_\text{I}=45^\circ$.}
		\label{Fig:res_b}
	\end{center}
\end{figure}

Figure~\ref{Fig:res_a} shows the numerical simulation results for the forward oblique wave incidence to the slab, shown in Fig.~\ref{Fig:bs_sch}, with $\epsilon_\text{r}=1$, $\delta_{\epsilon}=0.2$, $\gamma=1.2$, $d=3\lambda_0=3\times2\pi/k_0$, $\theta_\text{I}=45^\circ$ and $\omega_\text{0}=3$~GHz. It may be seen from this figure that an efficient beam splitting with significant transmission gain is achieved in the forward direction. Figures~\ref{Fig:res_c} and~\ref{Fig:res_d} provide the results for the oblique wave incidence from the right side and top, respectively, corresponding to $\theta_\text{I}=45^\circ$ and $\theta_\text{I}=-45^\circ$. The presented analytical and numerical results demonstrate that the dynamic beam splitter provides a perfect nonreciprocal beam splitting, in the lack of beam tilting. Moreover, it may be seen that, in contrast to conventional passive beam splitters, the beam splitting is achieved for a non-collimated beam. Other interesting features may be presented by changing the modulation parameters ($\gamma$, $\theta_\text{I}$ and $\epsilon_\text{av}$), including tunable transmission angles, unequal splitting ratio and unequal angles of transmission. Figure~\ref{Fig:res_b} compares the analytical and numerical results for the spectrum of the incident and transmitted electric fields in Fig.~\ref{Fig:res_a}. This figure shows that $3$dB transmission gain is achieved for each of transmitted beams in the forward excitation. Moreover, it may be seen from Fig.~\ref{Fig:res_b} that the undesired higher order harmonics, at $\omega=2m\omega_0$, are sufficiently weak so that the beam splitter safely operates at single frequency $\omega_0$.


\section{Closed form Solution for Electromagnetic Fields}\label{sec:closed}
It is shown in Sec.~\ref{sec:num} that by proper design of the band structure, a pure unidirectional beam splitting can be achieved in a obliquely illuminated STM slab. The analytical solution of the electromagnetic fields based on the double Bloch-Floquet decomposition of electromagnetic fields, presented in Sec.~\ref{sec:analyt}, provides an accurate solution for the fields scattered by such a slab, which is very useful. However, such an analytical solution does not provide a deep insight into the wave propagation inside the slab. In particular, it is of great interest to have an intuitive explanation about the effect of different parameters, e.g. $\delta_{\epsilon}$, $\gamma$, $k_x$ and $k_z$, on the wave propagation and energy exchange between the incident field $m=0$ and the excited first lower harmonic $m=-1$. Moreover, the accurate analytical solution, based on the mathematical modeling presented in Sec.~\ref{sec:analyt}, is achieved through a substantial computational cost. To resolve this issue, here we provide an approximate closed form solution for the electromagnetic fields propagating inside and transmitted from the STM beam splitter, which provides a clear picture of the transition between the incident and the first lower STHs.

As we showed in the previous section, given the weak transition of energy and momentum from the fundamental STH $m=0$ to higher order STHS except $m=-1$, the electric field inside the STM slab may be represented based on the superposition of the aforementioned two STHs, i.e., $m=0$ and $m=-1$. The electric field is then defined by
\begin{align}\label{eqa:1}
\begin{split}
E_\text{S}(x,z,t)=&a_{0}(z) e^{-i \left(k_x x+k_z z -\omega_0 t \right)}\\&+a_{-1}(z) e^{i \left(-k_x x+(q -k_z)z -\omega_0 t \right)},
\end{split}
\end{align}
where $a_{0}(z)$ and $a_{-1}(z)$ are the unknown field coefficients. We shall stress that, here the field coefficients are $z$-dependent since they include both the amplitude and the change in the spatial frequency (wavenumber) introduced by the ST modulation. Following the procedure provided in the supplemental material in Ref.~\cite{Taravati_PRB_NRBS_2018}, we insert the electric fields in~\eqref{eqa:1} into the wave equations in~\eqref{eqa:wave_eq}, and achieve a coupled differential equation for the field coefficients, i.e., 
\begin{subequations}
\begin{gather}\label{eqa:coupl}
\frac{d }{d z}  \begin{bmatrix} a_{0}(z) \\ a_{-1}(z) \end{bmatrix}
=
\begin{bmatrix}
M_0 & C_0 \\ C_{-1}& M_{-1}
\end{bmatrix}
\begin{bmatrix} a_{0}(z) \\ a_{-1}(z) \end{bmatrix},
\end{gather}
where
\begin{align}
M_0&=\frac{i k_0^2}{2k_z} (\epsilon_\text{av}-\epsilon_\text{r} ),\nonumber \\
M_{-1}&=\frac{ik_0^2}{2(k_z-q)} 
\left[\epsilon_\text{av}-\epsilon_\text{r}\frac{k_x^2+(q-k_z)^2}{k_0^2}  \right],\nonumber \\
C_0&=i\frac{\delta k_0^2}{4k_z}, \nonumber \\
C_{-1}&=i\frac{\delta k_0^2}{4(k_z-q)}.
\end{align}
\end{subequations}

The solution to the coupled differential equation in~\eqref{eqa:coupl} is given by~\cite{Taravati_PRB_NRBS_2018}
\begin{subequations}\label{eqa:coup_sol}
\begin{align}
a_{0}(z)=\frac{E_0}{2 \varDelta} &\bigg( (M_0-M_{-1}+\varDelta ) e^{\frac{M_0+M_{-1}+\varDelta}{2}z}  \\ \nonumber
&- (M_0-M_{-1}-\varDelta) e^{\frac{M_0+M_{-1}-\varDelta}{2}z}  \bigg),
\end{align}
\begin{align}
a_{-1}(z)=\frac{E_0 C_{-1}}{ \varDelta} \left( e^{\frac{M_0+M_{-1}+\varDelta}{2}z}  - e^{\frac{M_0+M_{-1}-\varDelta}{2}z}  \right),
\end{align}
\end{subequations}
where $\varDelta=\sqrt{(M_0-M_{-1})^{2}+4 C_0C_{-1}}$. For a given ST modulation ratio $\gamma$, the field coefficients in Eq.~\eqref{eqa:coup_sol} acquire different forms. In general, ST modulation is classified into three categories, i.e., subluminal ($0<\gamma<1$ or $v_\text{m}<v_\text{b}$), luminal ($\gamma\rightarrow 1$ or $v_\text{m}\rightarrow v_\text{b}$), and superluminal ($\gamma>1$ or $v_\text{m}>v_\text{b}$). 
\subsection{Subluminal and Superluminal ST Modulations}

Considering $\epsilon_\text{av}=\epsilon_\text{r}$, the $a_{0}(z)$ and $a_{-1}(z)$ in Eq.~\eqref{eqa:coup_sol} would be a periodic sinusoidal function with respect to $z$, if $\varDelta=\sqrt{(M_0-M_{-1})^{2}+4 C_0C_{-1}}$ is imaginary, i.e., $(M_0-M_{-1})^{2}+4 C_0C_{-1}<0$. By solving this, we achieve an interval for the luminal ST modulation, that is, 
\begin{equation}\label{eqa:sonic}
\gamma_\text{sub} <\frac{1 }{\sqrt{\epsilon_\text{av} +\delta_\epsilon }}
\leq\gamma_\text{lum}\leq
\frac{1 }{\sqrt{\epsilon_\text{av} -\delta_\epsilon }}<\gamma_\text{sup},
\end{equation}
where $\gamma_\text{sub}$, $\gamma_\text{lum}$ and $\gamma_\text{sup}$ are ST velocity ratio for subluminal, luminal and superluminal ST modulations, respectively. The interval for luminal ST modulation is called sonic regime in analogy with sonic boom effect in acoustics, where an airplane travels with the same speed or faster than the speed of sound. It should be noted that the luminal ST modulation interval in Eq.~\eqref{eqa:sonic} is exactly same as the one achieved from the exact analytical solution~\cite{Oliner_PIEEE_1963,Taravati_PRB_2017,Taravati_PRAp_2018}.\\

\begin{figure}
	\begin{center}
			\subfigure[]{\label{Fig:forw_sup}
				\includegraphics[width=1\linewidth]{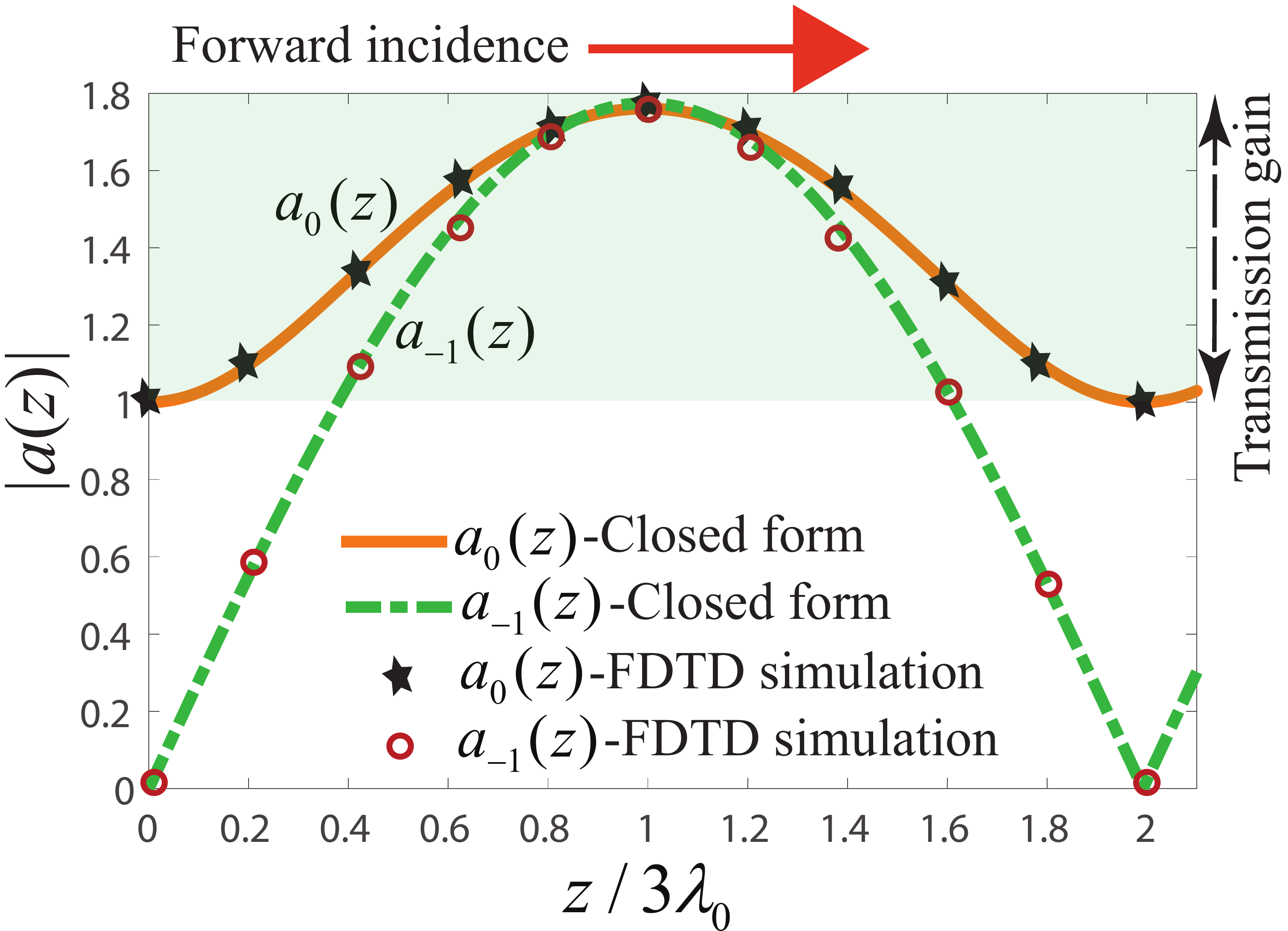}  } 
			\subfigure[]{\label{Fig:backw_sup}
				\includegraphics[width=0.95\linewidth]{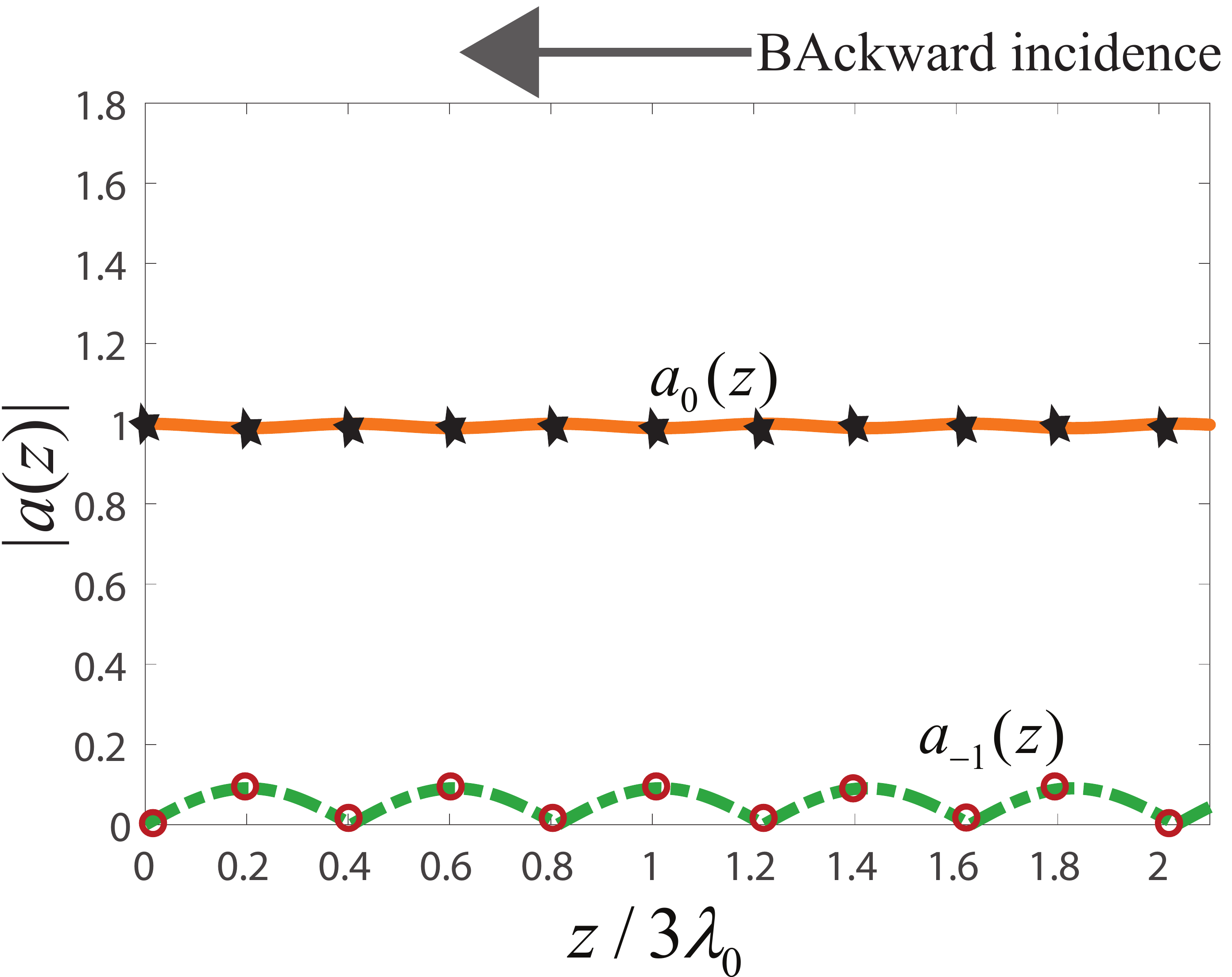}  } 
		\caption{Closed-form solution results and the FDTD numerical simulation results for the $z$-dependent absolute field coefficients in Eq.~\eqref{eqa:1}, i.e., $a_{0}(z)$ and $a_{-1}(z)$, inside the superluminal STM beam splitter, with $\gamma=1.2$ and $\delta_\epsilon=0.28$. (a)~Forward wave incidence, where the wave propagates from left to right. (b)~Backward wave incidence, where the wave propagates from right to left.}
		\label{Fig:sup}
	\end{center}
\end{figure}

Figure~\ref{Fig:forw_sup} plots the closed form and FDTD numerical simulation results for the absolute electric field coefficient inside the slab, with the wave incidence from the left side (forward incidence), considering superluminal ST modulation of $\gamma=1.2$ and $\delta_\epsilon=0.28$. It is seen from this figure that both $a_{0}(z)$ and $a_{-1}(z)$ possess periodic sinusoidal form and exhibit a substantial transmission gain at $z=3 \lambda_0$. Such a transmission gain may be tuned through the variation of $\gamma$ and $\delta_\epsilon$. This result is consistent with the transmission gain achieved in the FDTD numerical simulation results in Figs.~\ref{Fig:res_a} and~\ref{Fig:res_b}. The coherence length $l_\text{c}$, where both $a_{0}(z)$ and $a_{-1}(z)$ acquire their maximum amplitude is found as~\cite{Taravati_PRB_NRBS_2018}
\begin{align}\label{eq:iii}
l_\text{c}=\pi \left( \left[ \frac{ k_0^2[\epsilon_\text{av}-\epsilon_\text{r}]/k_z-q }{(\gamma-2)/(\gamma-1)}    \right]^2 +\frac{\delta^2 k_0^4}{4k_z (k_z-q)}  \right)^{-1}.
\end{align}\\

Figure~\ref{Fig:backw_sup} plots the result for the superluminal STM slab in Fig.~\ref{Fig:forw_sup}, except for wave incidence from the right side (backward incidence). It may be seen from this figure that, in contrast to the forward wave incidence where a substantial exchange of the energy and momentum between the $m=0$ and $m=-1$ STHs are achieved, here the incident wave passes through the slab with negligible alteration and minor transition of energy and momentum to the $m=-1$ ST harmonic. This is obviously in agreement with the nonreciprocal response presented in Figs.~\ref{Fig:res_a},~\ref{Fig:res_c} and~\ref{Fig:res_d}.
\subsection{Luminal ST Modulation}
It may be shown that the for the luminal ST modulation, where $\gamma\rightarrow1$, the field coefficients in Eq.~\eqref{eqa:coup_sol}, $a_{0}(z)$ and $a_{-1}(z)$, acquire pure real (or complex) forms. This yields exponential growth of the electric field amplitude along the STM slab. Hence, considering $\gamma=1$, the total electric field inside the STM slab reads
\begin{align}\label{eqa}
E_\text{S}(x,z,t)&|_{\gamma=1}=E_0 \cosh\left(\frac{\delta k_0^2}{4k_z}z\right) e^{-i \left(k_x x+k_z z -\omega_0 t \right)}\\ \nonumber &-i \frac{\delta k_0^2}{2 k_z}E_0 \sinh\left(\frac{\delta k_0^2}{4k_z}z\right) e^{i \left(-k_x x+(q -k_z)z -\omega_0 t \right)}.
\end{align}

\begin{figure}
	\begin{center}
		\subfigure[]{\label{Fig:forw_lum}
			\includegraphics[width=1\linewidth]{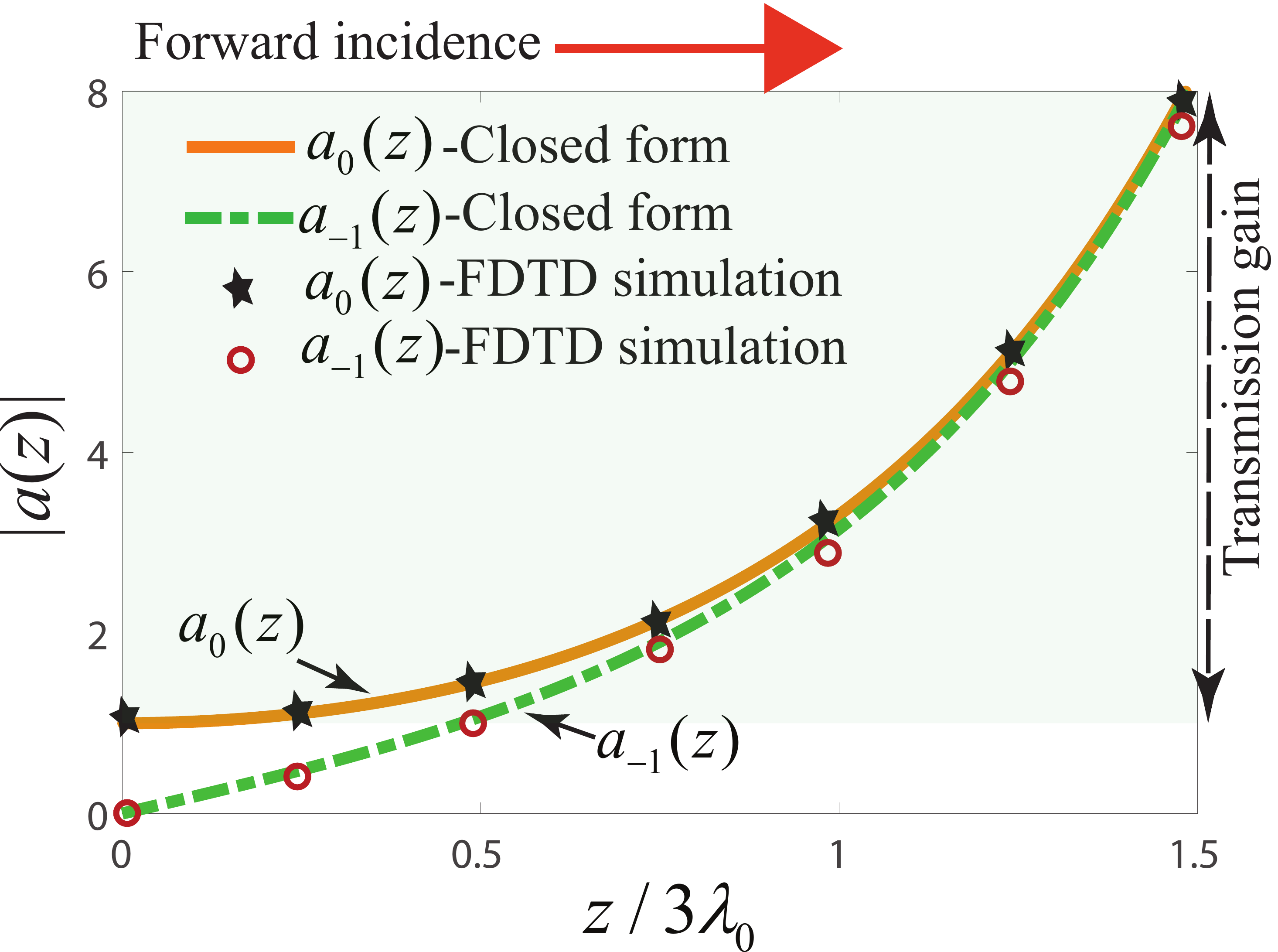}  } 
		\subfigure[]{\label{Fig:backw_lum}
			\includegraphics[width=1\linewidth]{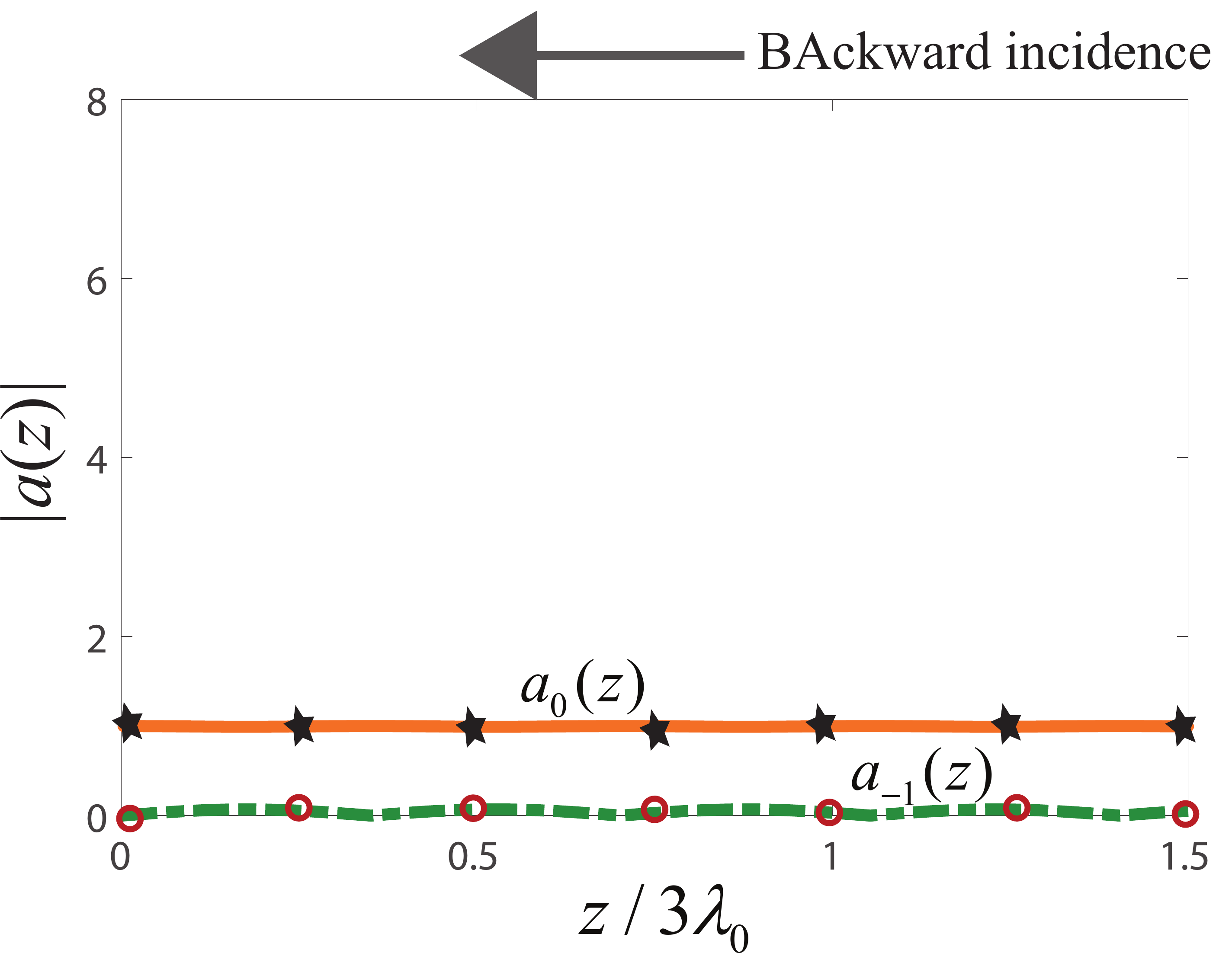}  } 
		\caption{Closed-form solution results and the FDTD numerical simulation results for the $z$-dependent absolute field coefficients in Eq.~\eqref{eqa:1}, i.e., $a_{0}(z)$ and $a_{-1}(z)$, inside the luminal STM beam splitter, with $\gamma=1$ and $\delta_\epsilon=0.28$. (a)~Forward wave incidence, where the wave propagates from left to right. (b)~Backward wave incidence, where the wave propagates from right to left.}
		\label{Fig:lum}
	\end{center}
\end{figure}

Figure~\ref{Fig:forw_lum} plots the closed form and FDTD numerical simulation results for the absolute value of the electric field coefficients $a_{0}(z)$ and $a_{-1}(z)$ inside the luminal ($\gamma=1$ and $\delta_\epsilon=0.28$) STM slab for forward wave incidence. It may be seen from this figure that both $a_{0}(z)$ and $a_{-1}(z)$ possess a non-periodic exponentially growing profile and exhibit a substantial transmission gain at $z\geq3 \lambda_0$. It should be noted that, the solution for the field coefficients presented in Eqs.~\eqref{eqa:coup_sol} and~\eqref{eqa} are very useful and provide a deep insight into the wave propagation inside the STM slab, especially for the luminal ST modulation (sonic regime), where the Bloch-Floquet-based analytical solution does not exist since the solution does not converge~\cite{Oliner_PIEEE_1963,Taravati_PRB_2017,Taravati_PRAp_2018}.

Figure~\ref{Fig:backw_lum} plots the result for the luminal STM slab in Fig.~\ref{Fig:forw_lum}, except for wave incidence from the right side (backward incidence). It may be seen from this figure that, in contrast to the forward wave incidence, here the incident wave passes through the slab with negligible alteration and minor transition of energy and momentum to the $m=-1$ ST harmonic.

\section{Discussion on Practical realization of Superluminal ST Modulation}\label{sec:prac}
To practically realize superluminal ST modulation (here $\gamma=1.2$), the phase velocity of the modulation should be greater than the velocity of the incident wave in the background (unmodulated) medium and not the velocity of light in vacuum~\cite{Cassedy_PIEEE_1967}. Considering a glass as the background medium with permittivity$>1.5$, achieving $\gamma=1.2$ would be realistic. For instance, one may use coupled structures with two different lines (possessing different phase velocities) for the input wave and the modulation~\cite{okwit1961constant,Taravati_LWA_2017,Taravati_PRB_SB_2017}. In such structures a modulation velocity greater than at least one of the characteristic velocities involved is required. In general, the way of achieving the fast pumping depends on the frequency range, as follows. 
\begin{itemize}
	\item At low frequencies, one may use filter constants which are appropriately selected for two weakly coupled transmission lines~\cite{okwit1961constant,Taravati_LWA_2017,Taravati_PRB_2017,Taravati_PRB_SB_2017}, one for the pump and one for the main incident wave.
	\item At ultra-high frequencies, a serpentine transmission line msupports the propagation of the main incident wave~\cite{okwit1962uhf}, which lowers the phase velocity relative to the modulation velocity.
	\item At microwave frequencies the pump wave is supported in a closed waveguide~\cite{hsu1960backward}, thereby achieving a fast phase velocity. 
\end{itemize}
\indent In addition, recently, there has been an experimental demonstration of time-modulated structure~\cite{reyes2016electromagnetic}, where the medium is periodically modulated in time only, representing the limiting case of an infinite modulation velocity, i.e., $q=0$ and hence $v_\text{m}=\Omega/q\rightarrow \infty$.

\section{Conclusion}\label{sec:conc}
We have introduced a unidirectional beam splitter and amplifier based on asymmetric coherent photonic transitions in obliquely illuminated space-time-modulated (STM) media. The operation of this dynamic beam splitter is demonstrated by both the analytical, closed-form and numerical simulation results. While the normally illuminated STM media have been previously used for the realization of various components, including insulators, parametric amplifiers and nonreciprocal frequency generators, this paper presented the first study investigating the oblique illumination of STM media. Accordingly, this paper proposed a forward-looking application of such dynamic media. The proposed unidirectional beam splitter is endowed with unique functionalities, including adjustable one-way transmission gain, tunable splitting angle and arbitrary unequal splitting power ratio, as well as high isolation, and hence, is expected to substantially reduce the source and isolation requirements of communication systems.
\bibliography{Taravati_Reference}

\pagebreak

\onecolumngrid
\vspace{1cm}
\textbf{SUPPLEMENTAL MATERIAL}
\\
	\renewcommand{\theequation}{S\arabic{equation}}
\renewcommand{\thesubsection}{\arabic{subsection}}
\renewcommand{\thesubsubsection}{3.\arabic{subsubsection}}
\setcounter{equation}{0}  
\setcounter{subsection}{0}  

\headsep = 40pt

The beam splitter is placed between $z=0$ and $z=d$, and represented by the space-time-varying permittivity of
\begin{equation}\label{eqa:permit}
\epsilon(z,t)=\epsilon_\text{av} + \delta_{\epsilon} \sin(q z-\Omega t),
\end{equation}
where $\Omega=2\omega_\text{0}$ and 
\begin{equation}
q= \frac{2 k_z}{\gamma}.
\end{equation} 

The electric field inside the beam splitter is defined based on the superposition of the $m=0$ and $m=-1$ space-time harmonics fields, i.e.,
\begin{align}\label{eqa:el}
E_\text{S}(x,z,t)=a_{0}(z) e^{-i \left(k_x x+k_z z -\omega_0 t \right)}+a_{-1}(z) e^{i \left(-k_x x+(q -k_z)z -\omega_0 t \right)},
\end{align}
and the corresponding wave equation reads
\begin{align}\label{eqa:wave_eq}
\frac{\partial^{2} \textbf{E}}{\partial x^{2}}+\frac{\partial^{2} \textbf{E}}{\partial z^{2}}= \frac{1}{c^2} \frac{\partial^{2} [\epsilon_\text{eq}(t,z) \textbf{E}]}{\partial t^{2}}.
\end{align}

Inserting the electric field in~\eqref{eqa:el} into the wave equation in~\eqref{eqa:wave_eq} results in 
\begin{equation}
\begin{split}
&\left( \frac{\partial^{2} }{\partial x^{2}}+\frac{\partial^{2} }{\partial z^{2}}  \right) \left[  a_{0}(z) e^{-i \left(k_x x+k_z z -\omega_0 t \right)}+a_{-1}(z) e^{i \left(-k_x x+(q -k_z)z -\omega_0 t \right)}   \right]\\
&   = \frac{1}{c^2} \frac{\partial^{2} }{\partial t^{2}} 
\left(\bigg[\epsilon_\text{av} + \frac{\delta}{2} e^{i(q z-2\omega_\text{0} t)}  + \frac{\delta}{2} e^{-i(q z-2\omega_\text{0} t)} \right] \left(a_{0}(z) e^{-i \left(k_x x+k_z z -\omega_0 t \right)}+a_{-1}(z) e^{i \left(-k_x x+(q -k_z)z -\omega_0 t \right)} \right) \bigg),
\end{split}
\end{equation}
and applying the space and time derivatives, while using a slowly varying
envelope approximation, yields
\begin{equation}\label{eqa:eq1}
\begin{split}
& \left[(k_x^2+k_z^2 )a_{0}(z) -2ik_z \frac{\partial a_{0}(z)}{\partial z} \right] e^{-i \left(k_x x+k_z z -\omega_0 t \right)}\\
&\qquad+\left[(k_x^2+(q-k_z)^2 )a_{-1}(z) -2i(k_z-q) \frac{\partial a_{-1}(z)}{\partial z} \right] e^{i \left(-k_x x+(q -k_z)z -\omega_0 t \right)}\\
& \qquad\qquad \qquad\qquad = \frac{\omega_0^2}{c^2} 
\left(\bigg[\epsilon_\text{av}+ \frac{\delta}{2} e^{i(q z-2\omega_0 t)}  + 9\frac{\delta}{2} e^{-i(q z-2\omega_0 t)} \right] a_{0}(z) e^{-i \left(k_x x+k_z z -\omega_0 t \right)}\\
&\qquad\qquad \qquad\qquad\qquad +\left[\epsilon_\text{av} + 9\frac{\delta}{2}  e^{i(q z-2\omega_0 t)}  + \frac{\delta}{2}  e^{-i(q z-2\omega_0 t)} \right] a_{-1}(z) e^{i \left(-k_x x+(q -k_z)z -\omega_0 t \right)}  \bigg),
\end{split}
\end{equation}

We then multiply both sides of Eq.~\eqref{eqa:eq1} with $e^{i \left(k_x x+k_z z -\omega_0 t \right)}$, which gives
\begin{equation}\label{eqa:eq2}
\begin{split}
& \left[(k_x^2+k_z^2 )a_{0}(z) -2ik_z \frac{\partial a_{0}(z)}{\partial z} \right] +\left[(k_x^2+(q-k_z)^2 )a_{-1}(z) -2i(k_z-q) \frac{\partial a_{-1}(z)}{\partial z} \right]  e^{i \left(q z -2\omega_0 t \right)}\\
&   = \frac{\omega_0^2}{c^2} 
\left(\bigg[\epsilon_\text{av}+ \frac{\delta}{2} e^{i(q z-2\omega_0 t)}  + 9\frac{\delta}{2} e^{-i(q z-2\omega_0 t)} \right] a_{0}(z)  +\left[\epsilon_\text{av} e^{i(q z-2\omega_0 t)}+ 9\frac{\delta}{2}  e^{i2(q z-2\omega_0 t)}  + \frac{\delta}{2}   \right] a_{-1}(z)  \bigg),
\end{split}
\end{equation}
and next, applying $\int_{0}^{\frac{\pi}{\omega_0}}dt$ to both sides of~\eqref{eqa:eq2} yields
\begin{equation}
\frac{d a_{0}(z)}{d z} 
= \frac{ik_0^2}{2k_z} 
\left( \left[\epsilon_\text{av}-\epsilon_\text{r} \right] a_{0}(z) +\frac{\delta}{2}    a_{-1}(z)  \right),
\end{equation}
which may be cast as
\begin{equation}\label{eqa:coup1}
\frac{d a_{0}(z)}{d z} = M_0 a_{0}(z)  +C_0    a_{-1}(z),
\end{equation}
where
\begin{equation}
M_0=\frac{ik_0^2}{2k_z} (\epsilon_\text{av}-\epsilon_\text{r} ), \qquad
C_0=\frac{i\delta k_0^2}{4k_z}.     
\end{equation}

Following the same procedure, we next multiply both sides of~\eqref{eqa:eq2} with $e^{-i \left(q z -2\omega_0 t \right)}$, and applying $\int_{0}^{\frac{\pi}{\omega_0}}dt$ in both sides of the resultant, which reduces to
\begin{equation}
\left[(k_x^2+(q-k_z)^2 )a_{-1}(z) -2i(k_z-q) \frac{\partial a_{-1}(z)}{\partial z} \right] 
= \frac{\omega_0^2}{c^2} 
\left( \frac{\delta}{2}    a_{0}(z)  +\epsilon_\text{av}  a_{-1}(z)  \right),
\end{equation}
which may be cast as
\begin{equation}\label{eqa:coup2}
\frac{d a_{-1}(z)}{d z}= C_{-1} a_{0}(z)  +M_{-1}    a_{-1}(z), 
\end{equation}
where 
\begin{equation}
M_{-1}=\frac{ik_0^2}{2(k_z-q)} 
\left[\epsilon_\text{av}-\epsilon_\text{r}\frac{k_x^2+(q-k_z)^2}{k_0^2}  \right], \qquad
C_{-1}=\frac{i\delta k_0^2}{4(k_z-q)}     
\end{equation}

Equations~\eqref{eqa:coup1} and~\eqref{eqa:coup2} form a matrix differential equation. We then look for independent differential equations for $a_0(z)$ and $a_{-1}(z)$, which is expressed by
\begin{align}\label{eqa:eq3}
\frac{d^{2}a_{0}(z)}{dz^{2}}-(M_0+M_{-1})\frac{da_{0}(z)}{dz}+(M_0M_{-1}-C_0C_{-1})a_{0}(z)=0,\\
\frac{d^{2}a_{-1}(z)}{dz^{2}}-(M_0+M_{-1})\frac{da_{-1}(z)}{dz}+(M_0M_{-1}-C_0C_{-1})a_{-1}(z)=0.
\end{align}
which are second order differential equations. Using the initial conditions of $a_{0}(0)=E_0$ and $a_{-1}(0)=0$ gives
\begin{equation}
a_{0}(z)=\frac{E_0}{2 \varDelta} \left( (M_0-M_{-1}+\varDelta ) e^{\frac{M_0+M_{-1}+\varDelta}{2}z}  - (M_0-M_{-1}-\varDelta ) e^{\frac{M_0+M_{-1}-\varDelta}{2}z}  \right),
\end{equation}
\begin{equation}
a_{-1}(z)=C_{-1}\frac{E_0 }{ \varDelta} \left( e^{\frac{M_0+M_{-1}+\varDelta}{2}z}  - e^{\frac{M_0+M_{-1}-\varDelta}{2}z}  \right),
\end{equation}
where
\begin{align}\label{eqa}
\varDelta=\sqrt{(M_0+M_{-1})^{2}-4 (M_0M_{-1}-C_0C_{-1})}=\sqrt{(M_0-M_{-1})^{2}+4 C_0C_{-1}}.
\end{align}

Assuming an imaginary result for $\varDelta$ (which occurs for subluminal and superluminal space-time modulations), $a_{0}(z)$ and $a_{-1}(z)$ acquire a periodic sinusoidal form, where the maximum amplitude of them occur at the coherence length $z=l_\text{c}$, where
\begin{equation}
\frac{d}{dz}a_{0}(z)|_{z=l_\text{c}}=\frac{d}{dz}a_{-1}(z)|_{z=l_\text{c}}=0,
\end{equation}
which corresponds to
\begin{align}\label{eq:iii}
l_\text{c}=\pi \left( \left[ \frac{ k_0^2[\epsilon_\text{av}-\epsilon_\text{r}]/k_z-q }{(k_z-q)/(k_z-q/2)}    \right]^2 +\frac{\delta^2 k_0^4}{4k_z (k_z-q)}  \right)^{-1}.
\end{align}

For luminal space-time modulation, where $\gamma\rightarrow 1$ and $q\rightarrow2k_z$, $a_{0}(z)$ and $a_{-1}(z)$ acquire exponentially growing profile, and hence, the total electric field inside the slab reads
\begin{align}\label{eqa}
E(x,z,t)=E_0 \cosh\left(\frac{\delta k_0^2}{4k_z}z\right) e^{-i \left(k_x x+k_z z -\omega_0 t \right)}-i \frac{\delta k_0^2}{2 k_z}E_0 \sinh\left(\frac{\delta k_0^2}{4k_z}z\right) e^{i \left(-k_x x+(q -k_z)z -\omega_0 t \right)},
\end{align}
which demonstrates the wave amplification of both $m=0$ and $m=-1$ ST harmonics, for the luminal space-time modulation.

\end{document}